\pdfoutput=1 
\documentclass[a4paper]{article}

\usepackage[pages=all, color=black, position={current page.south}, placement=bottom, scale=1, opacity=1, vshift=5mm]{background}
\SetBgContents{
	\tt This work is shared under a \href{https://creativecommons.org/licenses/by-sa/4.0/}{CC BY-SA 4.0 license} unless otherwise noted
}      

\usepackage[margin=1in]{geometry} 

\usepackage{amsmath}
\usepackage{amsthm}
\usepackage{amssymb}

\usepackage[utf8]{inputenc}
\usepackage{hyperref}

\usepackage[sort&compress,numbers,square]{natbib}

\theoremstyle{plain}

\theoremstyle{definition}

\usepackage{graphicx, color}
\graphicspath{{fig/}}

\usepackage{algorithm, algpseudocode} 
\usepackage{mathrsfs} 

\usepackage{slashbox,booktabs}

\usepackage{pdflscape}

\newcounter{Lcount}
\newcommand{\numsquishlist}{
   \begin{list}{\arabic{Lcount}. }
    { \usecounter{Lcount}
 \setlength{\itemsep}{-.1ex}      \setlength{\parsep}{0ex}
      \setlength{\topsep}{0ex}       \setlength{\partopsep}{0ex}
      \setlength{\leftmargin}{1em} \setlength{\labelwidth}{1em}
      \setlength{\labelsep}{0.1em} } }
\newcommand{\numsquishend}{\end{list}}

\newcommand{\squishlist}{
   \begin{list}{$\bullet$}
    { \setlength{\itemsep}{-.1ex}      \setlength{\parsep}{0ex}
      \setlength{\topsep}{0ex}       \setlength{\partopsep}{0ex}
      \setlength{\leftmargin}{.8em} \setlength{\labelwidth}{1em}
      \setlength{\labelsep}{0.5em} } }
\newcommand{\squishend}{\end{list}}

\definecolor{HLcolor}{rgb}{0,0.0,0}

\title{Modeling ICD-10 Morbidity and Multidimensional Poverty as a Spatial Network: Evidence from Thailand}
 \author{Pratana Kukieattikool \and Kittiya Ku-kiattikun \and Anukool Noymai \and Navaporn Surasvadi \and Jantakarn Makma \and Pubodin Pornratchpum \and Watcharakon Noothong \and Chainarong Amornbunchornvej\thanks{Corresponding author, email: chainarong.amo@nectec.or.th}}

\date{
	$^1$National Science and Technology Development Agency\\111 Phahonyothin Road, Khlong Nueng, Pathum Thani, 12120, Thailand \\%
}

\begin{document}
\maketitle


\begin{abstract}
Health and poverty in Thailand exhibit pronounced geographic structuring, yet the extent to which they operate as interconnected regional systems remains insufficiently understood. This study analyzes ICD–10 chapter–level morbidity and multidimensional poverty as outcomes embedded in a spatial interaction network. Interpreting Thailand’s 76 provinces as nodes within a fixed-degree regional graph, we apply tools from spatial econometrics and social network analysis—including Moran’s I, Local Indicators of Spatial Association (LISA), and Spatial Durbin Models (SDM)—to assess spatial dependence and cross-provincial spillovers.

Our findings reveal strong spatial clustering across multiple ICD–10 chapters, with persistent high–high morbidity zones—particularly for digestive, respiratory, musculoskeletal, and symptom-based diseases—emerging in well-defined regional belts. SDM estimates demonstrate that spillover effects from neighboring provinces frequently exceed the influence of local deprivation, especially for living-condition, health-access, accessibility, and poor-household indicators. These patterns are consistent with contagion and contextual influence processes well established in social network theory.

By framing morbidity and poverty as interdependent attributes on a spatial network, this study contributes to the growing literature on structural diffusion, health inequality, and regional vulnerability. The results highlight the importance of coordinated policy interventions across provincial boundaries and demonstrate how network-based modeling can uncover the spatial dynamics of health and deprivation.

\end{abstract}






\section{Introduction}
\label{sec:intro}
Digestive system diseases (ICD-10 Chapter XI) account for a substantial share of Thailand’s outpatient morbidity. Their distribution is shaped by structural determinants such as sanitation, food and water quality, and access to health services, all of which mirror long-standing regional socioeconomic inequalities. Recent work on spatial inequality and poverty in Thailand has shown that deprivation tends to form regional clusters and propagate through spatial dependence rather than appear as isolated pockets~\cite{gomez2024regional,sudswong2025occupational,puttanapong2022spatial}. This motivates treating ICD-10 morbidity not merely as province-specific outcomes but as \textbf{spatial phenomena embedded in regional socioeconomic systems}.

This paper examines the spatial structure of ICD-10 chapter–level morbidity across all 76 provinces of Thailand and its relationship to multidimensional poverty. We focus primarily on diseases of the digestive system (Labeled as C2) as a running example, and use several additional chapters (e.g. neoplasms, circulatory and respiratory diseases) to illustrate that the patterns observed for C2 are not idiosyncratic. Poverty is measured through the official Thai People Map and Analytics Platform (TPMAP), which decomposes deprivation into multiple dimensions (income, living conditions, health, education, accessibility, and poor-household count).

Our objectives are threefold:
\squishlist
\item To test whether ICD-10 morbidity ratios exhibit statistically significant spatial clustering at provincial level.
\item To assess how multidimensional poverty indicators align with these spatial patterns of morbidity.
\item To quantify spatial spillover effects—that is, the extent to which poverty in one province is associated with ICD-10 morbidity in its neighbors—using Spatial Durbin Models (SDM).
\squishend

We construct a 7-nearest-neighbors (KNN-7) spatial weights matrix over provinces, compute global Moran’s I and Local Indicators of Spatial Association (LISA) for ICD-10 chapters, and estimate SDMs in which both local and neighbor-lagged TPMAP indicators explain chapter-specific morbidity ratios. The results show strong spatial clustering of C2 and several other chapters, a prominent high-high cluster in Northeastern Thailand, and significant spillover effects from neighboring poverty. These findings reinforce a policy message: ICD-10 morbidity in Thailand is a regional, not purely provincial, phenomenon, and effective interventions require inter-provincial coordination.

\section{Related works}

\subsection{Socioeconomic Disadvantage as a Common Driver of Spatial Morbidity Patterns}

Health outcomes across many ICD–10 chapters—such as digestive, respiratory, genitourinary, and injury-related conditions—are shaped by shared structural determinants that disproportionately affect socioeconomically disadvantaged regions~\cite{phacharathonphakul2024socio,kaikeaw2023socioeconomic}. Digestive system diseases serve as a clear illustration of this mechanism: diarrhoeal diseases, intestinal infections, parasitic diseases, and several gastrointestinal cancers disproportionately affect populations exposed to inadequate sanitation, unsafe water, overcrowded housing, and poor nutrition \cite{PrussUstun2019, andres2021economic, WHO2016, wang2023global}. These conditions are themselves driven by structural poverty and limited public investment in basic infrastructure, especially in low- and middle-income countries \cite{barros2010socioeconomic, allen2017socioeconomic, wang2023global}. As a result, digestive disease burdens frequently concentrate geographically in disadvantaged communities, forming identifiable hot-spots where poor health and poverty reinforce one another over time \cite{wang2023global}. Such patterns closely align with broader evidence that poor living conditions and infrastructural deficits increase morbidity risks in low-income or underserved areas \cite{Krieger2002, WorldBank2017}.

Similar structural mechanisms operate across other high-burden ICD–10 chapters. Respiratory diseases are shaped by household air quality, environmental exposures, and crowding; genitourinary diseases arise from water contamination and sanitation deficits; and injury or poisoning patterns reflect occupational structures and access to emergency services. These categories, like digestive diseases, often form geographic clusters in settings where socioeconomic disadvantage and environmental vulnerabilities overlap \cite{Unger2007, Lilford2017}.  In short, deprivations jointly shape the distribution of morbidity, consistent with regional perspectives on inequality in Thailand~\cite{phacharathonphakul2024socio,kaikeaw2023socioeconomic}. 

Multidimensional poverty further compounds these risks. Income alone cannot explain vulnerability \cite{AMORNBUNCHORNVEJ2023e15947}; instead, deprivations in education, living standards, health access, transportation, and physical accessibility jointly constrain individuals’ exposure to environmental hazards, their mobility, and their ability to seek or sustain care. This resonates with regional and network-based perspectives on inequality in Thailand, where complex structures of occupational income, household vulnerability, and spatial dependence produce persistent regional disparities that simple summary indices like the Gini coefficient fail to capture \cite{gomez2024regional, sudswong2025occupational}.

Together, these structural and multidimensional determinants provide a foundation for interpreting the spatial clustering observed across multiple ICD–10 chapters in our analysis—not only digestive diseases but a wide array of conditions that co-vary with socioeconomic disadvantage.

\subsection{Spatial epidemiology and clustering of health outcomes}
Spatial epidemiology provides a set of tools for quantifying and visualizing how health outcomes cluster in space. Global Moran’s I~\cite{moran1948interpretation} measures overall spatial autocorrelation—whether high (or low) values tend to be located near other high (or low) values—while Local Indicators of Spatial Association (LISA)~\cite{anselin1995local} identify specific high-high, low-low, and outlier configurations. These methods have been widely used to \textbf{detect hot-spots of infectious and chronic diseases} and relate them to underlying socioeconomic and environmental conditions~\cite{jesri2021mapping}.

For instance, spatial analyses of intestinal infectious diseases in Latin America have documented significant positive Moran’s I statistics and LISA-identified high-high clusters overlapping with areas facing water scarcity, low educational attainment, and high rates of low-birth weight infants~\cite{lalangui2024space}. Similar approaches applied to national COVID-19 data revealed that high-mortality municipalities tended to be surrounded by other high-mortality, high-poverty municipalities, indicating that mortality risk was not randomly distributed but structured along socioeconomic gradients~\cite{birchenall2025exploring}. Studies of gastrointestinal cancers have shown that upper GI cancer mortality may cluster in the poorest urban neighborhoods, while lower GI cancers are more frequent in affluent areas, suggesting that \textbf{different digestive diseases respond to distinct poverty-related exposures}~\cite{reshadat2019comparative}.

\subsection{Spatial dependence, spillovers, and poverty traps}
A recurring finding in the health-poverty literature is that both deprivation and disease outcomes exhibit spatial dependence: poor or high-risk areas tend to be adjacent to other poor or high-risk areas~\cite{reshadat2019comparative}, forming “poverty belts” or “health belts” across administrative borders. At the same time, outcomes in a focal unit often depend not only on its own characteristics but also on the characteristics of its neighbors. Spatial econometric models—especially the Spatial Durbin Model (SDM)~\cite{koley2024use} and spatial lag models—were developed precisely to capture such \textbf{spillover effects}.

Empirical work has identified “poverty traps” in which clusters of counties or districts with persistent, multigenerational poverty experience significantly higher mortality from chronic liver diseases and liver cancers, even after controlling for individual demographic factors~\cite{ledenko2024poverty}. These areas exhibit strong spatial clustering and structural disadvantages that propagate across generations, implying that \textbf{location itself becomes a determinant of health}. Spatial Durbin estimates in such settings typically show that neighboring poverty indicators are significant predictors of local health outcomes, underscoring the importance of viewing territories as parts of an interconnected spatial network rather than as isolated entities.

\subsection{Spatial poverty and health in Thailand and Southeast Asia}
In Thailand, a growing body of work has documented spatial clustering of socioeconomic conditions, inequality, and poverty. Satellite-based and administrative data have been used to identify \textbf{regional inequality structures and poverty belts} using spatial statistics and econometrics~\cite{puttanapong2022spatial,puttanapong2022predicting}. These studies find that socioeconomic variables—such as income distribution, poverty incidence, and digital access—are not evenly distributed but form coherent regional patterns, with the Northeastern region consistently appearing as the most deprived.

Health research in Thailand and neighboring countries has begun to adopt similar spatial perspectives. Syndemic analyses have shown that parasitic liver fluke infection and leptospirosis co-occur and cluster in Northeastern Thailand, where poverty levels and agricultural occupational risks are highest, suggesting a \textbf{synergistic interaction between poverty and multiple health risks}~\cite{almanfaluthi2022burden}. Such findings reinforce the idea that health outcomes in Thailand cannot be detached from their spatial and socioeconomic context.

Despite this progress, there remains a gap in the literature: \textbf{no nationwide study has systematically analysed ICD-10 morbidity chapters as spatial processes linked to multidimensional poverty at provincial level}. Existing Thai work has focused either on specific diseases in specific regions or on spatial poverty without explicit linkage to a comprehensive health classification system. Our study contributes to this gap by combining ICD-10 chapter-level ratios with TPMAP multidimensional poverty indicators and applying spatial econometric models to quantify both local and neighbour-driven effects.

\section{Data}
\subsection{ICD-10 morbidity} 
We utilize publicly available data from the A-MED Care project\footnote{https://dashboard-amed-care.hii.in.th/}, collected in October 2025. The dataset contains 18,060,981 medical service transactions, comprising 11,571,595 records from patients visiting pharmacy shops and 6,489,386 records from patients receiving care at nursing centers.

We use aggregated outpatient morbidity data classified according to the International Statistical Classification of Diseases and Related Health Problems, 10th Revision (ICD-10), for all 76 provinces of Thailand. For each province and each ICD-10 chapter, we compute the ratio:

\begin{equation}
    ICD10_{ratio}{(c_j,i)}=\frac{c_j}{m_i}.
\end{equation}
Where $c_j$ is a number of cases in chapter $Cj$ in province $i$ and $m_i$ is a number of all ICD-10 cases in province $i$.

The primary focus of the empirical analysis is on diseases of the digestive system (C2 in our notation), but we also compute ratios for several other chapters (e.g. neoplasms, circulatory, respiratory) to examine whether the observed patterns generalize beyond C2.

\subsection{Multidimensional poverty indicators (TPMAP)}
Provincial-level poverty indicators are obtained from the Thai People Map and Analytics Platform (TPMAP)\footnote{www.tpmap.in.th
}, which offers a multidimensional assessment of household deprivation. The underlying survey data were collected in 2025 and were used to compute the Multidimensional Poverty Index (MPI)~\cite{alkire2021global}, the primary poverty metric adopted by the United Nations. For our analysis, we incorporate data from all 76 provinces of Thailand, excluding Bangkok. The specific TPMAP indicators used in this study are:

\begin{table}[ht]
\centering
\caption{Description of TPMAP Poverty Indicators}
\begin{tabular}{p{4cm} p{10cm}}
\hline
\textbf{Variable} & \textbf{Description} \\
\hline
\texttt{pov.rate} & Overall poverty rate. \\[4pt]

\texttt{CNT} & Count of poor households. \\[4pt]

\texttt{living} & Deprivation in living conditions (e.g., housing quality, water and sanitation). \\[4pt]

\texttt{health} & Deprivation in health-related aspects (access to health services, health status). \\[4pt]

\texttt{education} & Deprivation in education (attainment, schooling). \\[4pt]

\texttt{income} & Income deprivation. \\[4pt]

\texttt{accessibility} & Deprivation in physical accessibility (transport, distance to services). \\
\hline
\end{tabular}
\end{table}

These indicators mirror the multidimensional poverty structure in recent Thai spatial poverty work~\cite{gomez2024regional,AMORNBUNCHORNVEJ2023e15947}.

All ICD-10 and TPMAP variables are harmonized at provincial level and standardized where appropriate to facilitate model estimation and comparison.

\section{Methods}
\subsection{Spatial weights matrix}
We represent Thailand’s provinces as nodes in a spatial network and construct a 7-nearest-neighbors (7NN) spatial weights matrix $W$. For each province, the seven geographically nearest provinces (based on centroids) are treated as neighbors and assigned positive weights, row-standardized so that each row sums to one.
The choice of 7NN is motivated by two considerations:
\numsquishlist
	\item \textbf{Balanced neighborhood structure} – Thailand’s provinces differ markedly in shape and number of bordering neighbors. A purely contiguity-based matrix (e.g. queen contiguity) can yield some provinces with very few neighbors, leading to unstable local statistics and weight heterogeneity. KNN ensures that each province has the same number of neighbors, which improves the comparability of spatial effects and stabilizes estimation.
	\item \textbf{Consistency with previous Thai spatial work} – prior spatial analyses of poverty and inequality in Thailand have successfully used KNN-type or similar dense spatial structures to capture regional dependence~\cite{gomez2024regional}. In this work, $K=7$ provides a reasonable compromise between capturing local interactions and avoiding overly global connections.
\numsquishend
Sensitivity to alternative values of $K$ or contiguity definitions is discussed in the limitations.

\subsection{Exploratory spatial analysis}

We assess whether ICD--10 morbidity is spatially structured across Thailand's provincial
interaction network. Each province is treated as a node, and the spatial weights matrix $W$
defines its edges. Spatial dependence is examined using both global and local measures of
network autocorrelation.

\subsubsection{Global Moran's I and permutation test}

For each ICD--10 chapter ratio $y=ICD10_{ratio}$, we compute Global Moran's I:
\begin{equation}
I = 
\frac{n}{S_0}
\frac{
\sum_{i}\sum_{j \in \mathcal{N}_i} w_{ij}(y_i - \bar{y})(y_j - \bar{y})
}{
\sum_{i}(y_i - \bar{y})^2
},
\end{equation}
where $n$ is the number of provinces, $y_i$ is ICD--10 chapter ratio of province $i$, $\mathcal{N}_i$ is a set of neighbor provinces of $i$, $w_{ij}$ are row-standardized weights, 
$\bar{y}$ is the global mean, and $S_0 = \sum_i\sum_j w_{ij}$ is the sum of all weights. Note that since we use $7NN$,  $\forall j \in \mathcal{N}_i,w_{ij}=1/7$, otherwise $w_{ik}=0$ for any $k$ that is not a neighbor of $i$.
A significantly positive $I$ indicates that provinces with high (or low) morbidity
are connected to neighbors with similar values, implying network autocorrelation.

To evaluate significance, we conduct a Monte Carlo permutation test with $nsim$ random
permutations. For each simulation $k$:
\numsquishlist
    \item permute the morbidity vector to obtain $y^{(k)}$,
    \item compute $I^{(k)}$ using the same $W$,
    \item form the empirical reference distribution 
        $\{I^{(1)}, I^{(2)}, \ldots, I^{(nsim)}\}$.
\numsquishend
The pseudo--p-value is:
\begin{equation}
p = 
\frac{
1 + \sum_{k=1}^{nsim} \mathbf{1}\big( I^{(k)} \ge I_{\text{obs}} \big)
}{
1 + nsim
}.
\end{equation}
Where $\mathbf{1}(b)=1$ is the boolean $b$ is true, otherwise, it is zero. The permutation test holds the network structure $W$ fixed while randomizing node
attributes, providing a non-parametric test for spatial clustering.

\subsubsection{Local Moran's I (LISA)}

To identify where spatial dependence occurs, we compute Local Moran's I for each province:
\begin{equation}
I_i = 
\frac{(y_i - \bar{y})}{m_2}
\sum_j w_{ij}(y_j - \bar{y}),
\end{equation}
where
\begin{equation}
m_2 = \frac{1}{n} \sum_i (y_i - \bar{y})^2.
\end{equation}
$ I_i $ measures whether a node with high (or low) morbidity is surrounded by neighbors
with similar deviations from the mean.

Significance is again assessed using $nsim$ permutations:
\numsquishlist
    \item hold node $i$ fixed while permuting neighbor values,
    \item compute $I_i^{(k)}$ for each permutation,
    \item obtain a pseudo--p-value by comparing $I_i$ to 
    the empirical distribution $\{I_i^{(1)}, \ldots, I_i^{(nsim)}\}$.
\numsquishend

Each node is then classified into:
\begin{itemize}
    \item \textbf{High--High (HH)}: high morbidity with high-morbidity neighbors,
    \item \textbf{Low--Low (LL)}: low morbidity with low-morbidity neighbors,
    \item \textbf{High--Low (HL)}: a spatial outlier with high morbidity but low-morbidity neighbors,
    \item \textbf{Low--High (LH)}: a spatial outlier with low morbidity but high-morbidity neighbors.
\end{itemize}
A HH province must be satisfied 1) the province must have $ I_i $ is higher than other provinces significantly (determining by pseudo--p-value), and 2) the province and its neighbor's $z$ scores must be greater than 0. 
\begin{equation}
    z_i=\frac{(y_i-\bar{y})}{s_y}
\end{equation}
\begin{equation}
    z_{\text{lag}_i} = \sum_{j \in \mathcal{N}_i} w_{ij} \, z_j
\end{equation}
Where $s_y$ is a standard deviation of $y$. Specifically, for a HH province, $z_j>0, z_{\text{lag}_i}>0$. For a LL province, $z_j<0, z_{\text{lag}_i}<0$. The HL and LH provinces are typically treat as noise and can be determined in the second condition as $z_j>0, z_{\text{lag}_i}<0$ and $z_j<0, z_{\text{lag}_i}>0$ respectively. Other provinces that do not belong to any class above are labeled as insignificant. 

\subsubsection{Visualization}

Two mapping layers support interpretation:
\begin{enumerate}
    \item \textbf{Value maps}: show raw ICD--10 ratios across provinces (e.g.\ C2 morbidity share),
    revealing regional gradients in the provincial network.
    \item \textbf{LISA cluster maps}: show statistically significant HH, LL, HL, and LH provinces,
    highlighting interaction communities and structural outliers.
\end{enumerate}

This network-framed exploratory analysis provides evidence of spatial clustering and
motivates the use of spatial econometric models that incorporate node-to-node influence
through $W$.

\subsection{Spatial Durbin Model (SDM)}
To quantify both \textbf{local poverty effects} and \textbf{neighborhood spillovers}, we estimate a \textbf{Spatial Durbin Model} for each ICD-10 chapter. For a given chapter (e.g. C2), the model takes the form:

\begin{equation}
    y=\rho Wy+X\beta+WX\theta+\epsilon,
\end{equation}
where:
\squishlist
	\item $y$ is the $76\times 1$ vector of $ICD10_{ratio}$ values for the chapter.
	\item $X$ is the $76\times 7$ matrix of seven TPMAP poverty indicators.
	\item $Wy$ is the spatial lag of morbidity.
	\item $WX$ is the matrix of neighbor-lagged poverty indicators.
	\item $\rho$ is the spatial autoregressive coefficient capturing dependence in the outcome.
	\item $\beta$ are local (direct) effects of poverty.
	\item $\theta$ are spillover effects of neighboring poverty.
	\item $\epsilon$ is an error term.
\squishend
    
Models are estimated by maximum likelihood using standard spatial econometric routines. We use the function \textit{lagsarlm(type = "mixed")} in the R package \textit{spatialreg}~\cite{pebesma2023spatial}), which correspond to the SDM specification.

\subsection{Model evaluation and comparison}

For each chapter, we compare the SDM to a non-spatial linear model using:
\squishlist
	\item Log-likelihood and Akaike Information Criterion (AIC) – lower AIC indicates better fit.
	\item Significance of $\rho$– to test spatial dependence in the dependent variable.
	\item Significance of neighbor-lagged poverty coefficients $\theta$– to test spillover effects.
	\item LM tests for residual spatial autocorrelation – to ensure that spatial structure has been adequately captured.
\squishend
For C2, we also compare SDM specifications under alternative neighbor matrices (e.g. different K values) and find qualitatively similar patterns, reinforcing robustness (details omitted for brevity but discussed conceptually in the limitations).

\subsection{Network Perspective: Spatial Dependence as a Regional Interaction Network}
Although the analysis is presented in a spatial econometric framework, the underlying structure can be interpreted directly through a \textbf{network science perspective}. Provinces constitute the nodes of a regional interaction network, and the spatial weights matrix $W$ defines the edges linking each province to its nearest neighbors. Our use of a 7-nearest-neighbors (7NN) matrix corresponds to a fixed-degree network, where each node maintains the same number of ties. This reduces degree heterogeneity and stabilizes estimators in much the same way that fixed-degree social networks are used to study contagion dynamics or diffusion processes.

Interpreted this way, the Spatial Durbin Model (SDM) captures two mechanisms well-known in network analysis:
\numsquishlist
	\item Outcome contagion (spatial lag of y)
The term $\rho Wy$ represents a diffusion-like process: the morbidity ratio in a given province is shaped partly by the morbidity ratios of its neighbors. This mirrors behavioral or informational contagion in social networks, where the state of adjacent nodes influences the state of a focal node.
	\item Attribute spillover (spatially lagged covariates $WX$)
The neighbor-lagged TPMAP indicators represent indirect attribute effects, akin to contextual influence in social networks. The poverty characteristics of neighboring provinces (attributes of adjacent nodes) influence local health outcomes, capturing a structural vulnerability mechanism in which local conditions depend on the environments to which nodes are embedded.
\numsquishend

Viewing Thailand’s provinces as a \textbf{networked system} helps explain the persistence of spatial clusters in digestive diseases (C2) and other ICD-10 chapters. Provinces embedded in the same regional subnetwork share environmental exposures, food systems, water networks, occupational risks, and healthcare infrastructure. Thus, the high–high clusters in Northeastern Thailand correspond to \textbf{densely connected subnetworks} where poverty, environmental risks, and disease burdens reinforce one another.

This network interpretation also sharpens policy implications. Interventions targeting individual provinces (nodes) may be insufficient when the drivers of disease propagate through \textbf{network-level structures}. Instead, regional policies should target \textbf{communities or clusters} in the provincial network—akin to community detection in network science—where structural constraints and shared vulnerabilities create persistent health inequalities.

\section{Results}

\subsection{Global Moran’s I}
\label{sec:GlobMoran}
Global Moran’s I measures the overall degree of spatial autocorrelation in an outcome, indicating whether provinces with high (or low) morbidity tend to be located near provinces with similarly high (or low) values. Its p-value assesses the statistical significance of this spatial pattern—small p-values (e.g., < 0.05) confirm that the observed clustering is unlikely to occur by random chance. 

\begin{table}[htbp]
\centering
\small
\begin{tabular}{l l r r r}
\hline
ICD10 Cluster & ICD10 Name (Abbrev.) & Pop. Ratio & Global Moran's $I$ & $p$-value \\
\hline
C1  & Infectious \& parasitic diseases                     & 3.80\%  & 0.428  & 0.001 \\
C2  & Digestive system diseases                            & 10.43\% & 0.5533 & 0.001 \\
C3  & Eye \& adnexa diseases                               & 3.55\%  & 0.2519 & 0.001 \\
C4  & Genitourinary diseases                               & 0.66\%  & 0.3328 & 0.001 \\
C5  & Musculoskeletal \& connective tissue diseases        & 19.53\% & 0.5334 & 0.001 \\
C6  & Nervous system diseases                              & 3.34\%  & 0.1686 & 0.003 \\
C7  & Respiratory system diseases                          & 37.94\% & 0.6971 & 0.001 \\
C8  & Skin \& subcutaneous diseases                        & 3.15\%  & 0.3809 & 0.001 \\
C9  & Endocrine, nutritional \& metabolic diseases         & 1.86\%  & 0.5311 & 0.001 \\
C10 & Injury \& poisoning                                   & 2.24\%  & 0.4515 & 0.001 \\
C11 & Mental \& behavioural disorders                      & 0.31\%  & 0.1201 & 0.018 \\
C12 & Symptoms \& abnormal findings (NEC)                  & 12.33\% & 0.4367 & 0.001 \\
C13 & Ear \& mastoid diseases                              & 0.50\%  & 0.1979 & 0.001 \\
C14 & Blood \& immune disorders                            & 0.35\%  & 0.4251 & 0.001 \\
\hline
\end{tabular}
\caption{Population ratio and Global Moran's $I$ statistics for ICD-10 morbidity across provinces.}
\label{tb:GbMoran}
\end{table}

In Table~\ref{tb:GbMoran}, the population ratios (Pop. Ratio) indicate that healthcare utilization is highly uneven across ICD-10 categories, with respiratory diseases (C7, 37.94\%), musculoskeletal conditions (C5, 19.53\%), and digestive diseases (C2, 10.43\%) accounting for the majority of medical visits. These high-volume categories also exhibit some of the strongest spatial clustering, as reflected by their elevated Global Moran’s I values, suggesting that common medical burdens are not randomly distributed but form distinct regional hotspots. Even less prevalent categories, such as endocrine/metabolic diseases (C9) and symptoms/abnormal findings (C12), show significant spatial autocorrelation, highlighting the presence of shared structural or environmental factors shaping morbidity patterns across provinces. Overall, the combined patterns of population ratio and Moran’s I indicate that both the frequency of healthcare-seeking behaviors and their spatial structures are driven by underlying regional influences rather than isolated provincial characteristics.

\subsection{Spatial clustering of ICD-10 chapters}
\label{sec:SpatialClust}

To examine how morbidity patterns vary across Thailand, we analyzed four ICD-10 chapters with the largest population ratios that cover 80\% of population: C7 respiratory diseases (37.94\%), C5 musculoskeletal and connective tissue diseases (19.53\%), C12 symptoms and abnormal clinical findings (12.33\%), and C2 digestive system diseases (10.43\%). Their value maps reveal clear geographic variation in the proportion of visits attributable to each chapter, with certain provinces consistently exhibiting substantially higher burdens than others. These disparities are not random; the corresponding z-score maps show that provinces with above-average morbidity tend to cluster spatially, suggesting shared contextual, environmental, or behavioral influences.

\begin{figure}
    \centering
    \includegraphics[width=\textwidth,height=\textheight,keepaspectratio]{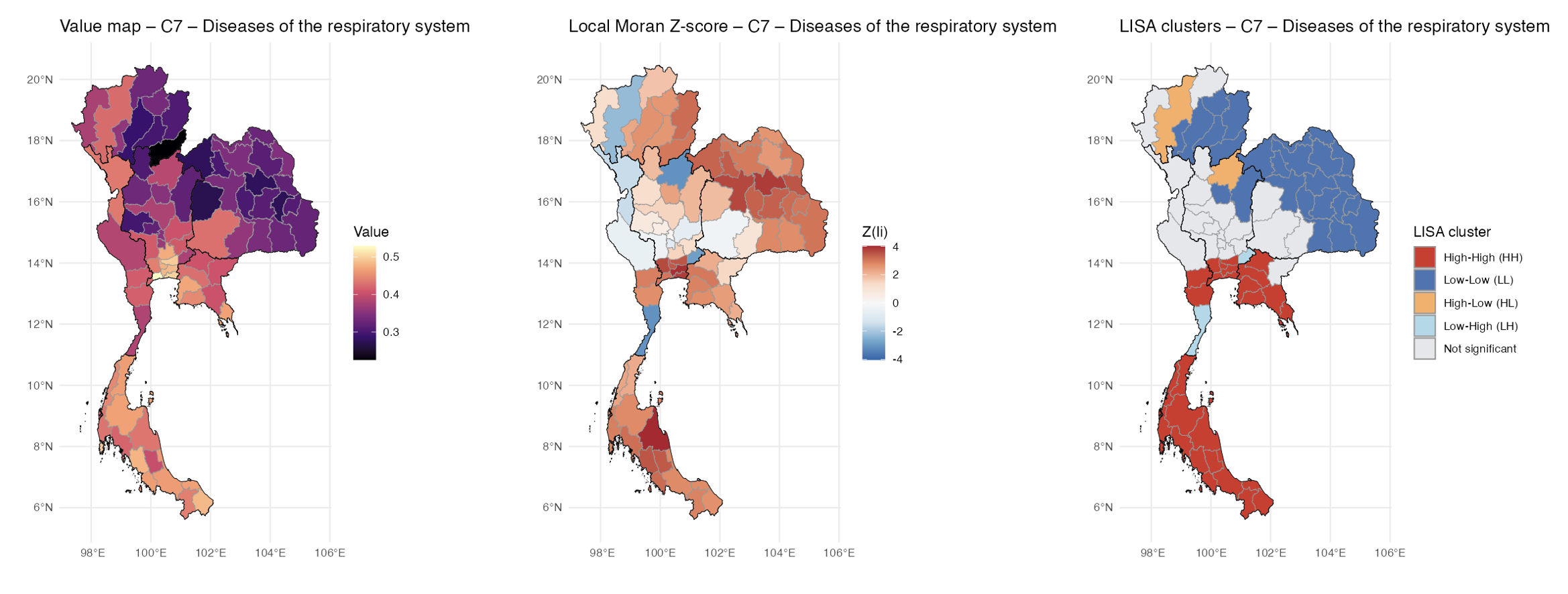}
    \caption{Maps of Respiratory system diseases (C7) (left) ICD-10 chapter ratios, (middle) Z-score of local Moran’s I, (right) LISA map of C7}
    \label{fig:C7maps}
\end{figure}

\begin{figure}
    \centering
    \includegraphics[width=\textwidth,height=\textheight,keepaspectratio]{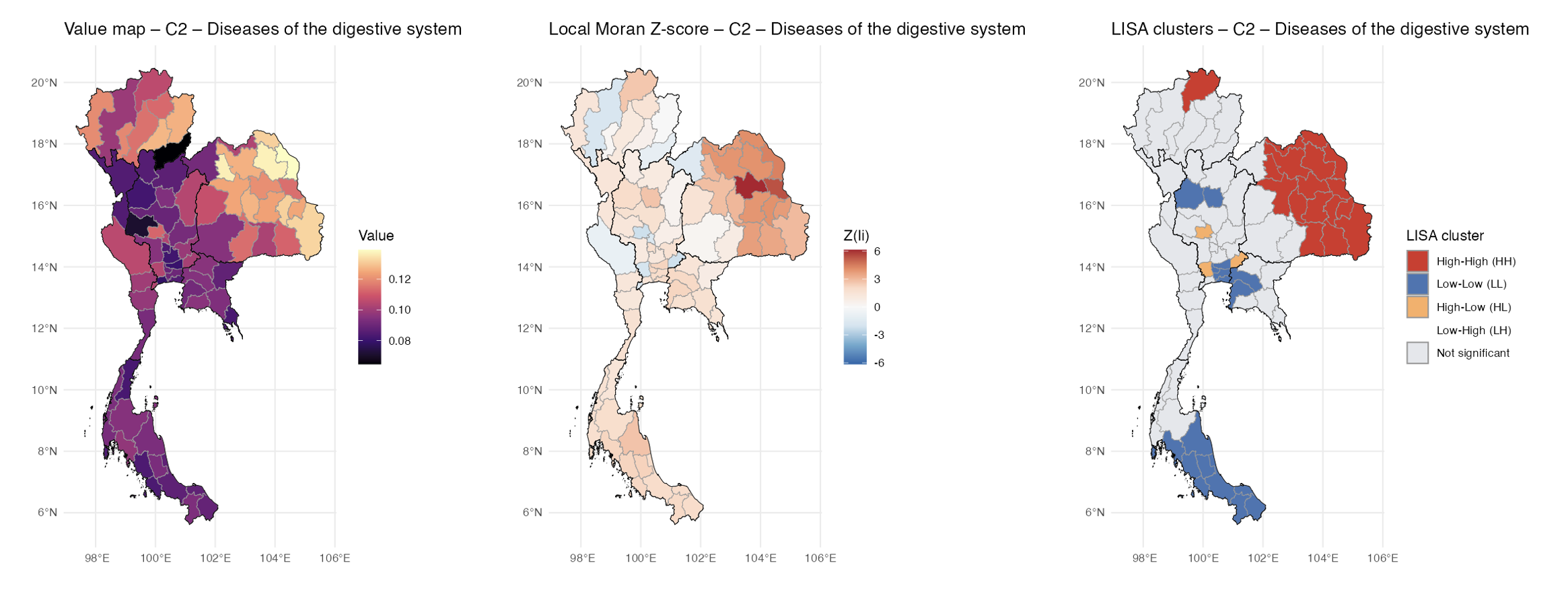}
    \caption{Maps of Diseases of the digestive system (C2) (left) ICD-10 chapter ratios, (middle) Z-score of local Moran’s I, (right) LISA map of C2}
    \label{fig:C2maps}
\end{figure}

\begin{figure}
    \centering
    \includegraphics[width=\textwidth,height=\textheight,keepaspectratio]{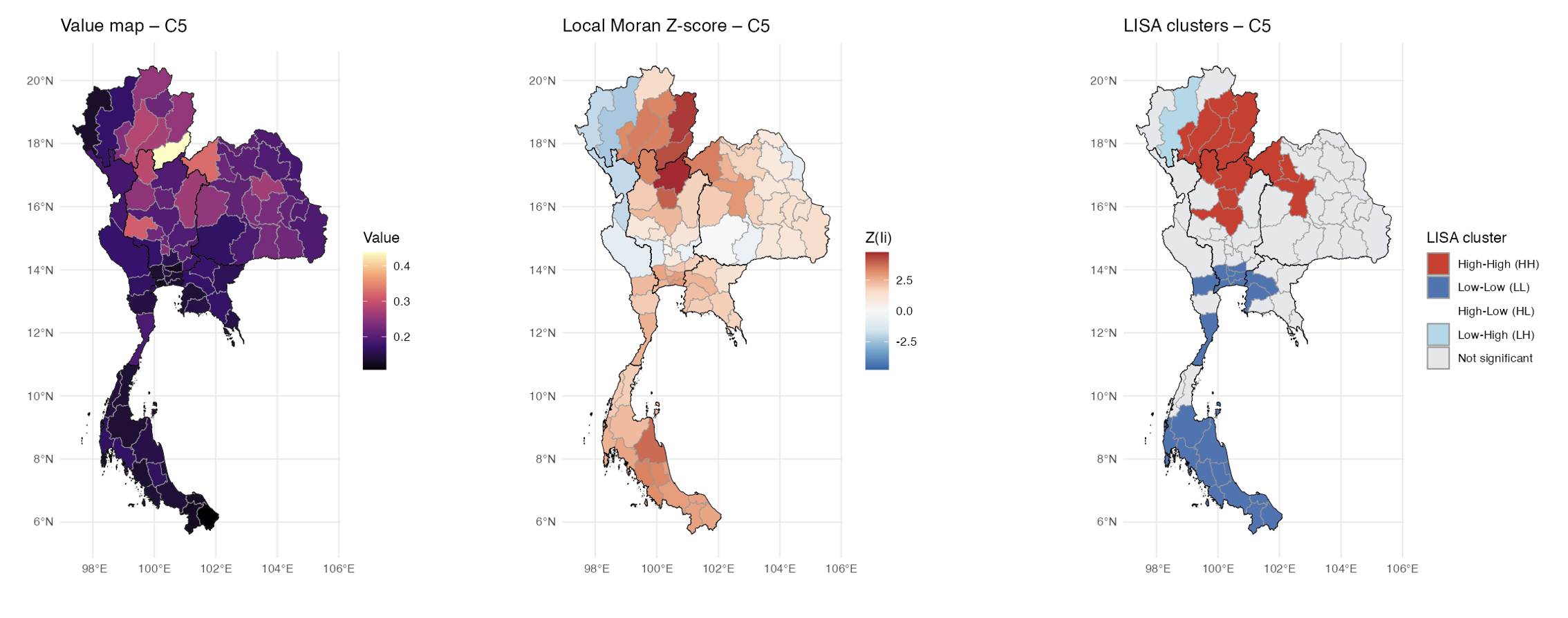}
    \caption{Maps of Musculoskeletal \& connective tissue diseases (C5) (left) ICD-10 chapter ratios, (middle) Z-score of local Moran’s I, (right) LISA map of C5}
    \label{fig:C5maps}
\end{figure}

\begin{figure}
    \centering
    \includegraphics[width=\textwidth,height=\textheight,keepaspectratio]{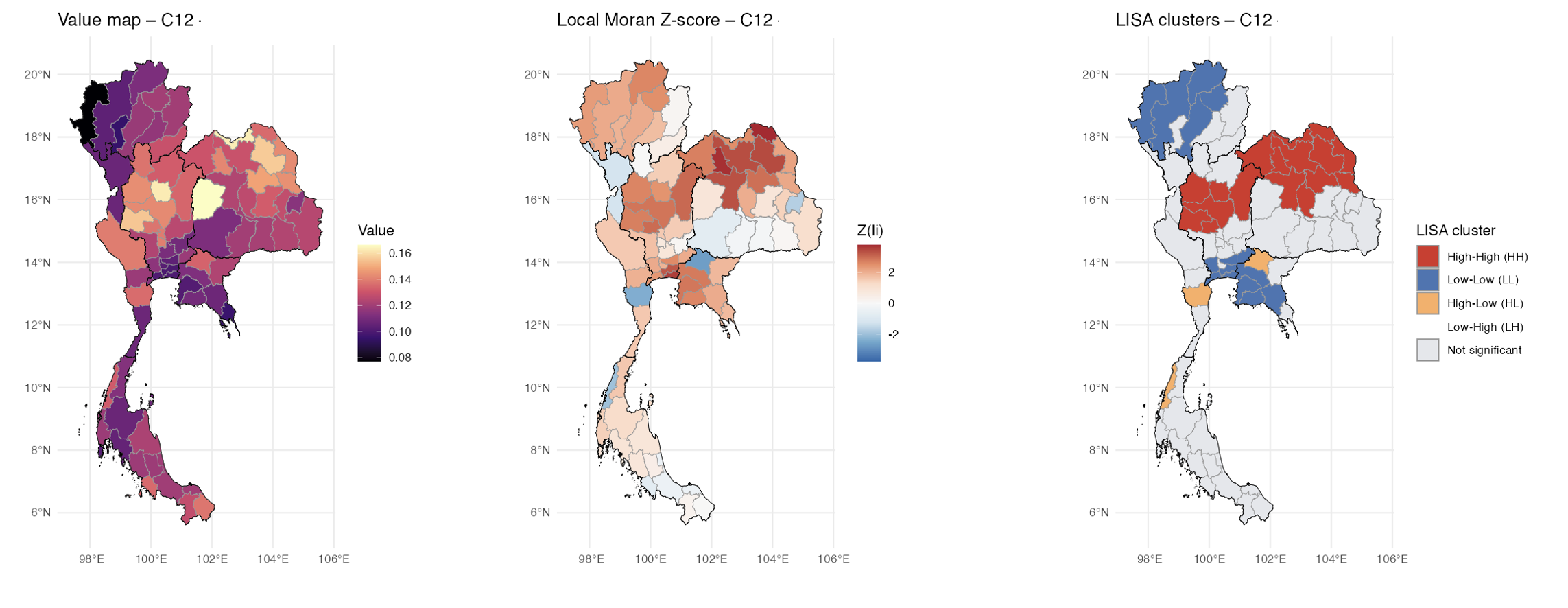}
    \caption{Maps of Symptoms \& abnormal findings (NEC) (C12) (left) ICD-10 chapter ratios, (middle) Z-score of local Moran’s I, (right) LISA map of C12}
    \label{fig:C12maps}
\end{figure}

Across all four chapters, the LISA cluster maps highlight statistically significant spatial structures. C7 in Fig~\ref{fig:C7maps} displays the strongest and broadest clustering, with extensive High–High (HH) regions in the western, eastern and deep southern provinces and Low–Low (LL) regions in the Northeast—consistent with its high Global Moran’s I value (0.6971) and reflecting regional environmental or seasonal drivers. C2 in Fig~\ref{fig:C2maps} shows a more sharply regionalized pattern, with a pronounced HH cluster in the Northeast and a clear LL cluster in the South, indicating localized digestive-disease risk factors. C5 in Fig~\ref{fig:C5maps} forms moderately strong clusters, with HH provinces in the North and upper Central region and LL clusters in the South, suggesting regionally shared demographic or occupational characteristics. C12 in Fig~\ref{fig:C12maps} exhibits the weakest but still significant clustering (Moran’s I = 0.4367), with smaller HH and LL pockets and several HL/LH outliers, consistent with its status as a broad, non-specific diagnostic category.

Taken together, these maps demonstrate that Thailand’s most common causes of healthcare utilization form distinct and interpretable spatial clustering patterns, with respiratory diseases showing the most intense spatial concentration, followed by musculoskeletal, digestive, and symptom-based presentations. The consistent emergence of HH and LL clusters underscores the importance of regional context—including environmental exposures, demographic structure, and healthcare-seeking behavior—in shaping provincial morbidity patterns.

\subsection{Spatial Durbin results}
\label{sec:SpaDurbRes}
\begin{table}[htbp]
\centering
\small
\begin{tabular}{lrrllrr}
\hline
ICD & $\rho$ & $p(\rho)$ & Significant direct effects & Significant indirect (lag) effects & AIC & LM test $p$-value \\
\hline
C1  & 0.293 & 0.139   & --        & health--                      & -508 & 0.471 \\
C2  & 0.351 & 0.0377  & living+   & health--, accessibility--, CNT+ & -444 & 0.742 \\
C3  & 0.405 & 0.0449  & --        & --                            & -544 & 0.397 \\
C4  & 0.321 & 0.105   & --        & health+, living+, CNT--      & -709 & 0.0881 \\
C5  & 0.683 & 2.02e-05& --        & --                            & -242 & 0.996 \\
C6  & -0.014& 0.955   & income+   & education--                   & -511 & 0.380 \\
C7  & 0.706 & 3.83e-07& CNT+      & health+, CNT--               & -248 & 0.995 \\
C8  & 0.179 & 0.384   & pov.rate+ & health--, living--, pov.rate+, CNT+ & -476 & 0.758 \\
C9  & 0.502 & 0.00529 & --        & --                            & -545 & 0.755 \\
C10 & 0.570 & 0.000916& --        & health+                       & -486 & 0.790 \\
C11 & 0.383 & 0.0800  & income+   & --                            & -711 & 0.0228 \\
C12 & 0.745 & 9.28e-08& --        & education--                   & -419 & 0.513 \\
C13 & -0.227& 0.385   & --        & living--, education+          & -789 & 0.0286 \\
C14 & 0.568 & 0.000669& --        & --                            & -618 & 0.865 \\
\hline
\end{tabular}
\caption{Summary of Spatial Durbin Model results for ICD-10 chapters (significant effects at $p<0.05$). ``+'' and ``--'' denote positive and negative coefficients, respectively.}
\label{tb:SDMres}
\end{table}

Across ICD-10 chapters, the Spatial Durbin Model results in Table~\ref{tb:SDMres} reveal substantial variation in how morbidity patterns relate to poverty dimensions and spatial spillovers. Spatial dependence ($\rho$) is strong and significant for several chapters—including C5, C7, C10, C12, and C14—indicating clear clustering where provinces resemble their neighbors in these morbidity burdens. Direct poverty effects are relatively sparse, with only a few chapters (C2, C6, C11) showing significant within-province associations, suggesting that most ICD categories are not driven strongly by local poverty conditions alone. In contrast, indirect (lag) effects are more common and more varied, particularly in C2, C4, C7, C8, C10, C12, and C13, demonstrating that neighboring provinces’ poverty profiles often exert measurable spillover influence on local morbidity. The direction of these effects differs by chapter, reflecting heterogeneous spatial mechanisms behind different disease categories. Collectively, these results imply that while some ICD chapters are locally linked to specific poverty dimensions, many morbidity patterns in Thailand are shaped more by regional spatial structures and cross-border poverty spillovers than by isolated provincial conditions.

\subsection{Cross-chapter patterns}
\label{sec:CroessPattRes}
\begin{figure}[htbp]
  \centering
  \includegraphics[width=1.0\textwidth]{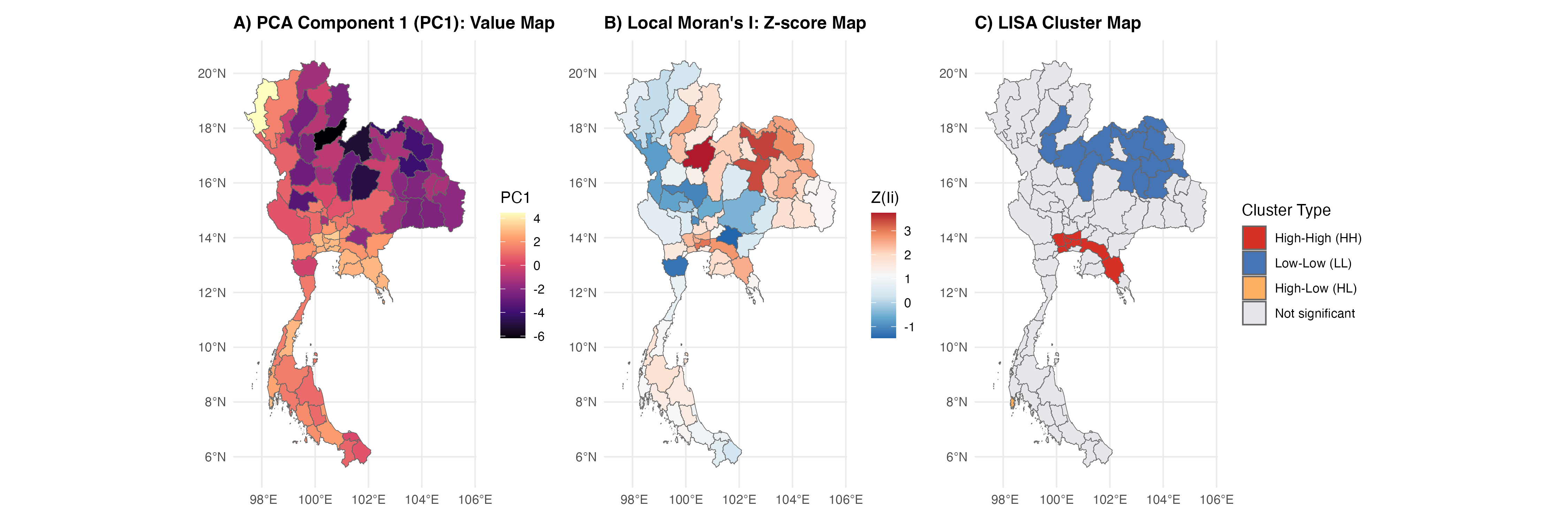}
  \caption{Spatial patterns of the first principal component (PC1) of ICD-10 morbidity.
(Left) PC1 value map showing the composite morbidity gradient.
(Middle) Local Moran’s z-scores indicating spatial autocorrelation.
(Right) LISA clusters identifying High–High regions in the North/Central area and Low–Low regions in the South.}
  \label{fig:PCA_LISA_3Panel}
\end{figure}

\begin{figure}[htbp]
  \centering
  \includegraphics[width=0.8\textwidth]{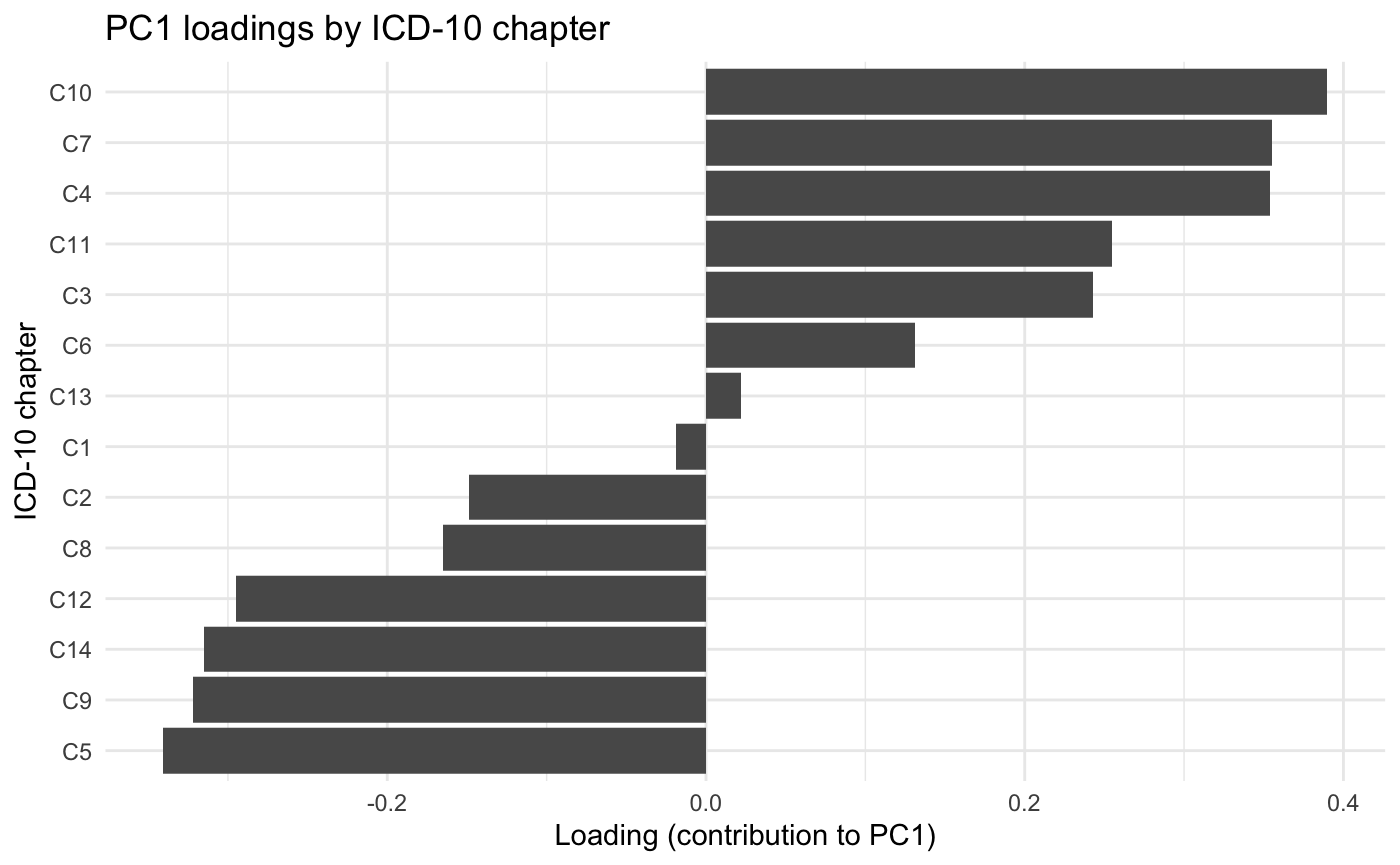}
  \caption{Contributions of each ICD-10 chapter to PC1. Chapters with large positive loadings (e.g., C10, C7, C4) drive higher composite morbidity in High–High regions, while chapters with negative loadings (e.g., C5, C9, C14) characterize lower-scoring regions.}
  \label{fig:PC1loadings}
\end{figure}

\begin{figure}[htbp]
  \centering
  \includegraphics[width=0.8\textwidth]{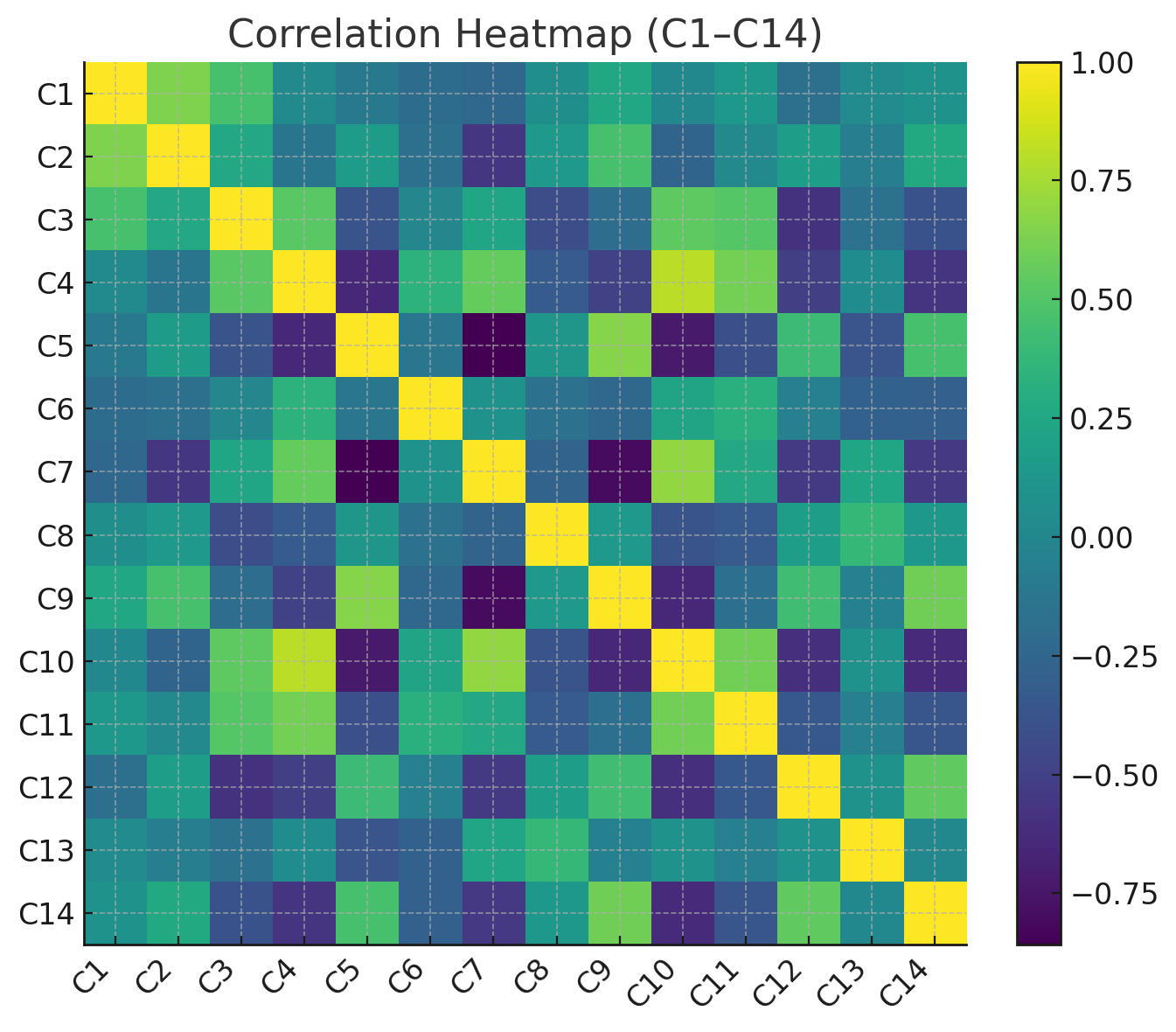}
  \caption{Pairwise correlations among ICD-10 chapter ratios (C1–C14). Positively correlated chapters cluster together, highlighting shared cross-chapter morbidity patterns that align with PCA results.}
  \label{fig:ICD10Corr}
\end{figure}

To examine how different ICD-10 chapters co-vary across Thailand, we applied Principal Component Analysis (PCA) to the full set of chapter-level morbidity ratios (C1–C14) and analyzed the resulting multivariate spatial pattern using Local Moran’s I. The first principal component (PC1) captured the dominant shared structure of morbidity across provinces. The PC1 value map in Fig.~\ref{fig:PC1loadings} shows a clear geographic gradient: provinces in the South display elevated composite morbidity levels, whereas provinces in the North and Northeast exhibit consistently lower scores. This spatial gradient is reinforced by the Local Moran’s I z-score map, which highlights strong positive spatial autocorrelation throughout the Southern half of the country, indicating that provinces with high composite morbidity values tend to cluster together rather than appear in isolation.

The LISA cluster map further highlights significant spatial structuring in this multivariate burden. A well-defined High–High (HH) cluster emerges in the Central ans eastern region, signaling a zone where multiple ICD-10 chapters simultaneously present high burdens. In contrast, the North and upper Central region forms a coherent Low–Low (LL) cluster, reflecting provinces with consistently low multichapter morbidity. These spatial patterns align closely with the PCA loadings in Fig~\ref{fig:PC1loadings}, where chapters such as C10 (injury/poisoning), C7 (respiratory diseases), and C4 (genitourinary diseases) contribute most strongly and positively to PC1, while chapters such as C5 (musculoskeletal diseases), C9 (endocrine/metabolic diseases), and C14 (blood and immune disorders) show strong negative loadings that distinguish the low-PC1 provinces.

The cross-chapter correlation heatmap in Fig.~\ref{fig:ICD10Corr} provides an important complementary perspective. It reveals several clusters of ICD-10 chapters that tend to rise and fall together across provinces. Notably, respiratory diseases (C7), injuries (C10), digestive diseases (C2), and genitourinary diseases (C4) share moderate to strong positive correlations, forming a cohesive block of high-burden chapters that also load positively on PC1. Meanwhile, chapters with negative PC1 loadings—such as musculoskeletal (C5), endocrine/metabolic (C9), and blood/immune disorders (C14)—are also moderately intercorrelated, but weakly correlated with the high-burden group. These patterns indicate that ICD-10 chapters do not vary independently across Thailand; instead, they form coherent morbidity clusters, with some categories consistently co-occurring in the same provinces due to shared environmental, demographic, or socioeconomic determinants.

Together, these results show that Thailand’s morbidity landscape contains robust multivariate regional syndromes, where multiple ICD-10 chapters rise and fall together in predictable geographic patterns. The alignment between PCA components, LISA clusters, and cross-chapter correlations suggests that underlying structural factors—rather than chapter-specific effects—shape the spatial distribution of health burdens across the country.
\section{Discussion}
The empirical results align with international evidence that health outcomes and multidimensional poverty tend to co-cluster geographically~\cite{lalangui2024space,birchenall2025exploring,reshadat2019comparative,almanfaluthi2022burden,ledenko2024poverty}. For several other ICD-10 chapters, we find that provincial morbidity ratios are not randomly distributed but exhibit strong spatial autocorrelation and clear HH and LL clusters. For example, the high-high C2 cluster in Northeastern Thailand corresponds to longstanding poverty belts identified in Thai spatial poverty research~\cite{gomez2024regional,sudswong2025occupational,puttanapong2022spatial,puttanapong2022predicting}, where deficits in living conditions, health access, and infrastructure are particularly acute.

The SDM estimates highlight that neighbor-lagged poverty indicators are often more important than local poverty indicators in explaining ICD10 chapter ratios. This is consistent with the notion of spatial spillovers and poverty traps: regional environments characterized by concentrated deprivation, inadequate sanitation, and limited health access shape disease risk across multiple adjacent provinces, regardless of minor differences in their local conditions. Our findings echo studies that find persistent liver disease mortality in multi-generational poverty clusters~\cite{ledenko2024poverty}, co-occurring infectious diseases in poor rural regions~\cite{almanfaluthi2022burden}, and synchronized COVID-19 mortality patterns in high-poverty municipalities~\cite{birchenall2025exploring}.
 
In the Thai context, this suggests that digestive disease epidemiology and other ICD10 diseases should be understood as a regional phenomenon embedded in the same spatial networks that shape inequality and poverty. The correspondence between ICD10 HH clusters and TPMAP-defined deprivation belts, plus the significance of neighbor-lagged poverty indicators, adds a health dimension to existing evidence that Thai socioeconomic structures are strongly spatially interdependent~\cite{gomez2024regional,sudswong2025occupational,puttanapong2022spatial}.

\section{Policy implications}


The findings of this study have several implications for public health and poverty alleviation policy in Thailand.
\numsquishlist
\item	\textbf{Region-based interventions are essential.}
 Statistically significant spatial clustering of ICD-10 morbidity across provinces (Global Moran’s I and LISA results in Sections~\ref{sec:GlobMoran}~\ref{sec:SpatialClust}; Table~\ref{tb:GbMoran} and Figures~\ref{fig:C7maps}-\ref{fig:C12maps}) indicates that disease burdens are structured at a regional level rather than confined within administrative boundaries. This suggests that province-specific interventions alone may be insufficient, and that region-based or inter-provincial approaches are more appropriate for addressing clustered morbidity. In this context, Thailand’s existing health region structure provides a relevant administrative framework through which spatially informed interventions could be operationalized.
\item	\textbf{Integrated poverty–health strategies are required.}
 The Spatial Durbin Model results show that multidimensional poverty is associated with morbidity through both local effects and spillover effects from neighboring provinces (Section~\ref{sec:SpaDurbRes}, Table~\ref{tb:SDMres}). In particular, indirect effects related to living conditions, health deprivation, and physical accessibility are statistically significant for several ICD-10 chapters. These findings highlight the importance of addressing broader structural conditions alongside healthcare delivery, and suggest that coordinated interventions at the health-region level may be more effective than isolated actions at the provincial level.
\item	\textbf{Coordination across neighboring provinces and regions is critical.}
 The prominence of neighbor-lagged poverty effects (Section~\ref{sec:SpaDurbRes}) implies that improvements within a single province may be constrained by persistent deprivation in surrounding areas. This underscores the need for coordination not only across provincial boundaries but also across health regions, particularly where spatial clusters of poverty and morbidity extend beyond formal administrative borders.
\item	\textbf{Spatially integrated monitoring can support policy targeting.}
 The consistency of spatial patterns across multiple ICD-10 chapters, together with the multivariate results from the Principal Component Analysis (Section~\ref{sec:CroessPattRes}; Figures~\ref{fig:PCA_LISA_3Panel}–\ref{fig:ICD10Corr}), suggests that health burdens tend to co-occur in predictable regional configurations. Integrating ICD-10 morbidity indicators with multidimensional poverty measures into spatial monitoring systems at the health-region level could support the identification and prioritization of structurally vulnerable areas and enhance evidence-based planning.
\numsquishend

Overall, these findings point to the need for a shift from predominantly province-centered policies toward spatially informed and regionally coordinated strategies that explicitly account for spatial clustering and cross-provincial spillovers identified in this study

\section{Limitations}
This study has several limitations that should be acknowledged.
First, the analysis is based on \textbf{cross-sectional, province-level data}. The results cannot be interpreted as individual-level causal effects of poverty on health, and there is a risk of ecological fallacy. Spatial associations between multidimensional poverty and ICD-10 morbidity may mask important intra-provincial heterogeneity in both socioeconomic conditions and disease burdens~\cite{gomez2024regional,puttanapong2022predicting,lalangui2024space,birchenall2025exploring,reshadat2019comparative,almanfaluthi2022burden}.

Second, both ICD-10 morbidity and TPMAP poverty indicators may be subject to \textbf{measurement error and reporting bias}. Diagnostic coding practices, health-seeking behaviors, and reporting completeness likely differ across provinces. Likewise, TPMAP indicators are derived from models and administrative data, which may under- or over-estimate certain dimensions of deprivation~\cite{gomez2024regional,sudswong2025occupational,puttanapong2022spatial}. These issues could attenuate or distort estimated associations.

Third, the spatial specification relies on a \textbf{KNN-7 weights matrix}. Although this choice is motivated by the need for balanced neighborhood structures and is consistent with previous Thai spatial analyses~\cite{gomez2024regional,sudswong2025occupational,puttanapong2022spatial}, alternative specifications (e.g. first-order queen contiguity, different K values, distance-based weights) may yield somewhat different estimates of spatial dependence. A systematic sensitivity analysis across multiple weight matrices would strengthen robustness but is beyond the scope of this paper.

Fourth, the SDMs control only for \textbf{multidimensional poverty dimensions} and do not include other relevant variables such as environmental exposures (water quality, land use), behavioral risk factors (diet, alcohol consumption, smoking), or detailed health-system characteristics (provider density, hospital quality). Omitted variables may partially confound or mediate the observed associations between poverty and digestive disease burden~\cite{lalangui2024space,birchenall2025exploring,reshadat2019comparative,almanfaluthi2022burden,ledenko2024poverty}.

Finally, our analysis focuses on \textbf{one period of aggregated morbidity} and does not model temporal dynamics. We therefore cannot examine how spatial clusters of ICD-10 morbidity and poverty evolve over time or how policy interventions might modify these patterns. Extending the framework to \textbf{spatio-temporal or Bayesian hierarchical models}—in line with recent spatial poverty work in Thailand~\cite{gomez2024regional}—would allow for a richer understanding of regional trajectories and causal mechanisms.

Despite these limitations, the consistency between the spatial clustering results, the SDM estimates, and prior literature on spatial health–poverty linkages suggests that the main substantive conclusions are robust at the provincial level.

In addition to the limitations discussed above, an important direction for future research is to examine how the data-driven spatial patterns identified in this study relate to Thailand’s existing health governance structure. While the present analysis models provinces as nodes in a spatial interaction network, health system planning and resource allocation in Thailand are formally organized through an official health region system administered by the Ministry of Public Health. Future work could therefore compare the spatial clusters and regional morbidity structures derived from ICD-10 morbidity and multidimensional poverty data with the boundaries of these official health regions, using quantitative measures such as cluster overlap, boundary-crossing rates, or partition similarity metrics. Such an analysis would help assess the degree of alignment between empirically observed health–poverty systems and existing administrative regions, and identify areas where cross-region coordination may be particularly important in addressing spatial spillovers.

\section{Conclusion}

This study reframes ICD-10 morbidity across Thai provinces as an outcome embedded in a spatial interaction network, rather than as a set of isolated provincial phenomena. By treating provinces as nodes within a fixed-degree spatial network and incorporating 7-nearest-neighbor interactions, we demonstrate that regional health outcomes—particularly for respiratory, digestive, musculoskeletal, and symptom-based disease categories—exhibit significant spatial autocorrelation and form consistent high-high and low-low clusters.

Our use of spatial econometric models, including the Spatial Durbin Model, captures two key mechanisms of network dependence: (i) \textit{outcome contagion}, where morbidity in one province reflects the morbidity levels of its neighbors; and (ii) \textit{attribute spillover}, where multidimensional poverty in neighboring provinces exerts measurable influence on local disease burden. These dynamics align with well-established principles in network science, such as diffusion, structural vulnerability, and contextual influence.

We find that neighbor-lagged poverty indicators often have greater explanatory power than local deprivation, highlighting the importance of structural interdependence. Additionally, multichapter principal component analysis reveals spatially coherent morbidity syndromes across the national network, emphasizing that health burdens co-evolve through shared regional exposures.

Taken together, these findings underscore the need for regionally coordinated interventions and spatially aware public health policy. Beyond Thailand, the framework we present is broadly applicable to other national settings where spatial data and administrative health records are available. This network-based approach contributes to the growing field of structural diffusion and spatial health inequality, and opens future research directions for dynamic, multiscale, and adaptive network modeling in epidemiology and social systems.

\bibliographystyle{apalike}


\end{document}